\documentstyle[aps]{revtex}
\draft
\begin{document}
\title{Maximum occupation number for composite boson states}

\author{S.~Rombouts, D.~Van Neck, K.~Peirs and L.~Pollet }
\address{Universiteit Gent, Vakgroep Subatomaire en Stralingsfysica
         \\
         Proeftuinstraat 86, B-9000 Gent, Belgium
         }
\maketitle

\begin{abstract}
One of the major differences between fermions and bosons is 
that fermionic states have a maximum occupation number of one, 
whereas the occupation number for bosonic states is in principle unlimited.
For bosons that are made up of fermions, 
one could ask the question to what extent the Pauli principle for the 
constituent fermions would limit the boson occupation number.
Intuitively one can expect the maximum occupation number to be
proportional to the available volume for the bosons
divided by the volume occupied by the fermions inside one boson,
though a rigorous derivation of this result has not been given before.
In this letter we show how the maximum occupation number can be calculated 
from the ground-state energy of a fermionic generalized pairing problem. 
A very accurate analytical estimate of this eigenvalue is derived.
From that a general expression is obtained 
for the maximum occupation number of a composite boson state,
based solely on the intrinsic fermionic structure of the bosons.  
The consequences for Bose-Einstein condensates of excitons in semiconductors
and ultra cold trapped atoms are discussed.
\end{abstract}

\section{Introduction}
Textbooks on quantum mechanics teach that fermionic states have a
maximum occupation number of one, whereas the occupation number of
bosonic states is unlimited. However, if the boson is actually a
many-fermion state (e.g. an atom), the Pauli principle results in a
maximum occupation number for the bosonic state,
irrespective of the nature of the boson-boson interactions. 
Though one can easily argue that the maximum occupation number for
a composite boson state should be of the order $V/r^D$, 
with $V$ the available volume, $D$ the dimension 
and $r$ the range of the intrinsic fermionic wave function 
of the boson~\cite{Hana77},
a rigorous and accurate derivation of this limit has not been given before.
Here we try to quantify this effect directly in  terms of the internal 
fermionic structure of the boson.
We find that the Pauli principle puts an upper limit 
for the occupation number of a composite boson state,
irrespective of the boson-boson interactions.
For a boson state composed of two fermions,
this {\em maximum occupation number} (MON) can be calculated as the
maximal eigenvalue of a fermionic generalized pairing operator.
Using variational principles we can derive a very accurate 
analytical estimate of this eigenvalue.
This allows us to write down analytical expressions for the MON
in a number of cases: for excitons in semiconductors,
for Gaussian wavefunctions and for atoms in Harmonic traps.

\section{The maximum occupation number and the generalized pairing problem}
As a first case, consider a number of bosons occupying 
a given one-boson state $\psi_0$.
The operator that creates one boson in the state $\psi_0$
can be written as combination of pairs of fermion creation operators:
$ \hat{b}^{\dagger}_0  
    = \sum_{i,j} B_{0, i j} \hat{a}^{\dagger}_{i} \hat{a}^{\dagger}_{j}$.
The number of bosons occupying the state $\psi_0$ can be evaluated as
the expectation value in the many-boson state
of the boson number operator $ \hat{N}_0 = \hat{b}^{\dagger}_0 \hat{b}_0$.
The MON is then given by the maximal eigenvalue 
of the number operator $  \hat{N}_0 $.
According to the Bloch-Messiah theorem 
we can always find a unitary transformation 
that brings the matrix $B_0$ in a canonical form:
\begin{equation}
 \hat{b}^{\dagger}_0  = \sum_{\alpha} c_{\alpha} \hat{a}^{\dagger}_{\alpha} 
                              \hat{a}^{\dagger}_{\bar{\alpha}},
\label{eq:bosedeco}
\end{equation}
where the indices $(\alpha,\bar{\alpha})$ correspond 
to conjugated pairs of canonical fermion states.
After this transformation, 
we see that the boson number operator is equivalent
to a fermionic generalized pairing operator
\begin{equation}
 \hat{N}_0 \equiv  \hat{G} = \sum_{\alpha, \alpha'} c^{*}_{\alpha}c_{\alpha'}
         \hat{a}^{\dagger}_{\alpha} \hat{a}^{\dagger}_{\bar{\alpha}} 
         \hat{a}_{\bar{\alpha'}} \hat{a}_{\alpha'}.
\label{eq:ghamilton}
\end{equation}
The MON corresponds to the ground-state energy 
of the pairing Hamiltonian $\hat{H}= -\hat{G}$.
This energy can be calculated efficiently 
using algebraic methods~\cite{Rich64,Feng98,Neck01},
for particle numbers up to a few thousand.
Here, instead, we  derive a very accurate 
and practical analytical estimate of the MON.
At the same time, this gives a simple and accurate analytical estimate 
of the ground-state energy of the pairing Hamiltonian $\hat{H}$.

\noindent {\em Upper bound:}
Consider the matrix representation of the operator $\hat{N}_0$ 
in the basis spanned by the states that are a product 
of $N$ pairs of fermions $(\alpha,\bar{\alpha})$.
The eigenstate corresponding to the largest eigenvalue 
is contained in this space~\cite{Burg96}.
The {\it Gershgorin circle theorem}~\cite{Golu89} 
results in the following upper bound:
\begin{equation}
 \mbox{MON} \leq \max_{\sigma} \left[
      \sum_{j=1}^N |c_{\sigma_j}|^2 + \left(\sum_{j=1}^N |c_{\sigma_j}|\right)
    \left(S - \sum_{j=1}^N |c_{\sigma_j}|\right) \right],
\label{eq:gershgorin}
\end{equation}
where $ S= \sum_{\alpha} |c_{\alpha}|$, 
$\sigma$  denotes a state of $N$ pairs 
distributed over the canonical pair states $(\sigma_1,\ldots,\sigma_N)$,
and the $c_{\sigma_j}$ denote the corresponding Bloch Messiah coefficients.
Because of the normalization of the fermion pair state of Eq.(\ref{eq:bosedeco}),
the first term in Eq.(\ref{eq:gershgorin}) is bounded by one,
\begin{equation}
\sum_{j=1}^N |c_{\sigma_j}|^2 \leq \sum_{\alpha} |c_{\alpha}|^2 =1.
\end{equation}
The second term is bounded because
\begin{equation}
x (S-x) \leq S^2/4 ,
\end{equation}
for any real number $x$,
in casue for $x=\sum_{j=1}^N |c_{\sigma_j}|$.
Using these facts we can obtain a slightly relaxed but more general bound, 
valid for any $N$:
\begin{equation}
  \mbox{MON} \leq \frac{S^2}{4} + 1.
\label{eq:ubound}
\end{equation}

\noindent {\em Lower bound:}
Taking the following fully paired state as a trial state,
\begin{equation}
 |\psi_T(N) \rangle = 
 \left( \sum_{\alpha}  \hat{a}^{\dagger}_{\alpha} 
           \hat{a}^{\dagger}_{\bar{\alpha}} \right)^N |0\rangle,
\end{equation}
where $|0\rangle$ denotes the zero-particle vacuum state,
we can derive a lower bound from the variational principle that
\begin{equation}
 \mbox{MON} \geq  \langle \psi_T | \hat{G} | \psi_T \rangle
     / \langle \psi_T  | \psi_T \rangle.
\end{equation}
This results in 
\begin{equation}
 \mbox{MON} \geq \frac{N \left[N-1+ (M-N) S^2 \right] }{M (M-1)} ,
\label{eq:lboundn}
\end{equation}
with $N$ the number of bosons (fermion pairs) in the trial state and
$M$ the number of canonical pair states
$ \hat{a}^{\dagger}_{\alpha} \hat{a}^{\dagger}_{\bar{\alpha}} |0\rangle$.
The maximum of the lower bound is found to be very close to the upper bound.
For $S \gg 1$, this maximum occurs at $N \simeq M/2$. 
Taking $N=(M+1)/2$ in Eq.(\ref{eq:lboundn}) leads to
\begin{equation}
 \mbox{MON} \geq \frac{1}{4} \left[S^2 +1 \right] (1+\frac{1}{M}).
\label{eq:lbound}
\end{equation}
For a realistic description we will have 
to take the limit of $M$ going to infinity,
because the Hilbert space has an infinite dimensionality.
This means that the total  number of fermion pairs can go to infinity too
(the quantity $N$ in Eq.(\ref{eq:lboundn}) ).
If the quantity $S$ has a finite value in this limit, 
then the upper bound in Eq.(\ref{eq:ubound}) is finite, 
and hence also the lower bound in Eq.(\ref{eq:lboundn}) will have a finite value.
Eq.(\ref{eq:ubound}) and Eq.(\ref{eq:lbound}) 
allow one to put general but very stringent limits on the MON:
\begin{equation}
 \frac{S^2}{4} + \frac{1}{4} \leq \mbox{MON} \leq \frac{S^2}{4} + 1.
  \label{eq:estimate}
\end{equation}
To estimate the MON we have to calculate the quantity $S$.
In two cases we can derive detailed expressions for this quantity:
in the case that the one-boson wave function corresponds to a plane wave
(for instance in a cubic volume $V$ with periodic boundary conditions) 
and in the case of a Gaussian one-boson and intrinsic wave function.

\section{The maximum occupation number for an exciton state}
For a boson in a cubic volume $V$ with periodic boundaries,
we can express the wave function in the momentum coordinates:
\begin{equation}
\psi({\bf k}_1,{\bf k}_2) \propto 
   \delta({\bf k}_1-{\bf k}_2) F\left[({\bf k}_1+{\bf k}_2)/2\right],
\end{equation}
where $F({\bf{k}})$ corresponds to the Fourier transform 
of the real space intrinsic wave function.
The canonical pair states of the Bloch-Messiah decomposition for such a state 
are given by the momentum pairs $({\bf k},-{\bf k})$,
\begin{equation}
 \hat{b}^{\dagger}_0  = \sum_{\bf k} F( {\bf k} ) \hat{a}^{\dagger}_{\bf k}  
                       \hat{a}^{\dagger}_{-{\bf k}}
\end{equation}
Then the quantity $S$ from Eq.(\ref{eq:estimate}) can be calculated as
\begin{equation}
 S = \sum_{\bf k} |F( {\bf k} )| .
\end{equation}

This expression can be applied directly to electron-hole excitons 
in semiconductors~\cite{Hana77,Rice77,Hens77,Keld68}.
Two recent letters~\cite{Comb01,Tang02} addressed the question 
to what extent these excitons can be regarded as bosons.
A criterion to assess that, was derived from the expectation values
$\langle \phi^{(N)} | \hat{B}^\dagger \hat{B} | \phi^{(N)} \rangle $ and
$ \langle \phi^{(N)} | 1- \left[\hat{B}, \hat{B}^\dagger\right] 
  | \phi^{(N)} \rangle$,
with $\hat{B}^{\dagger}$ the one-exciton creation operator and
\begin{equation}
  | \phi^{(N)} \rangle \propto \left(\hat{B}^{\dagger}\right)^N | 0 \rangle.
\end{equation}
They expect the boson picture to be valid up to 
\begin{equation}
 N \simeq \frac{V}{\lambda a_x^3},
\end{equation}
with $V$ the available volume, $a_x$ the exciton Bohr radius 
and $\lambda$ a scaling factor.
Their model allows them to propose a value $\lambda \simeq 100 $ up to $400$.
Compare this with a value of $\lambda \simeq 1$ that was suggested
from the disappearance of the exciton binding energy at the Mott density, 
due to Coulomb screening~\cite{Klin95}.
The state $|\phi^{(N)} \rangle$ can be seen as a number-projected 
Hartree-Fock-Bogoliubov state.
It can be expected that the true many-exciton state will differ significantly 
from this mean-field state at high exciton densities.
Whatever the true state might be, 
the occupation number $\langle \hat{B}^\dagger \hat{B} \rangle$
will be limited by the upper bound for the MON from Eq.(\ref{eq:ubound}).
Assuming that the intrinsic wave function of the excitons 
takes the form of a $1s$ hydrogen wave function,
we obtain the momentum representation
\begin{equation}
F({\bf k}) = \frac{8 \sqrt{\pi a_x^3/V }}{\left(1+k^2 a_x^2 \right)^2},
\end{equation}
with $a_x$ the exciton Bohr radius.
We find that $S^2=V / (\pi a_x^3)$, such that
\begin{equation}
\mbox{MON} = \frac{V}{4 \pi a_x^3}.
\label{eq:exmon}
\end{equation}
This means that the Pauli principle forbids 
having more than $V/(4 \pi a_x^3)$ excitons.
The corresponding scaling factor $\lambda$ is given by
\begin{equation}
 \lambda = 4 \pi.
\end{equation}
For the two-dimensional box with surface $V$, we have
\begin{equation}
F({\bf k}) = \frac{\sqrt{2\pi a_x^2/V}}{\left(1+k^2 a_x^2/4\right)^{3/2}},
\end{equation}
and hence MON $= V / (\frac{\pi}{2} a_x^2) $ and $ \lambda = \pi/2$.

\section{The maximum occupation number for a Gaussian state}
Another case where we can derive an expression for the quantity $S$,
is in the case of Gaussian wave functions 
for the one-boson state and for the intrinsic two-fermion state.
For that it is useful to work with a reduced density matrix, 
integrating out one of the fermion coordinates:
\begin{equation}
\rho \left({\bf r}_1,{\bf r}_1'\right) =
  \int \psi({\bf r}_1,{\bf r}_2)^{\dagger}
       \psi({\bf r}_1',{\bf r}_2) d {\bf r}_2.
\end{equation}
The values $c_{\alpha}$ of the Bloch-Messiah decomposition correspond to the
square roots of the eigenvalues of the matrix $\rho$.
So we can write $ S= \mbox{Tr}\left[\rho^{\frac{1}{2}}\right]$.
Assuming a Gaussian shape for the wave functions
we can write the reduced density matrix as
\begin{equation}
\rho \left({\bf r},{\bf r}'\right) \propto \ e^{-\frac{r^2+r'^2}{4 W^2}}
 \,  e^{-\frac{({\bf r}-{\bf r}')^2}{8 v^2}},
 \label{eq:rhogauss}
\end{equation}
where $W$ gives the width of the one-boson state, 
while the scale $v$ corresponds to 
the width of the fermion distribution inside one boson.
Note that the square of a matrix of this form has again the same form,
with different coefficients.
By equating these coefficients, 
we can find an explicit form of $\rho^{\frac{1}{2}}$,
and hence the trace,
\begin{equation}
 S= \left[\frac{W}{v} \left( 1 + \sqrt{1+\frac{v^2}{W^2}} \right) \right]^{3/2}
  \simeq \left[2   \frac{W}{v} \right]^{3/2}.
\end{equation}
This means that we can estimate the MON by
\begin{equation}
 \mbox{MON} \simeq  2 \left(\frac{W}{v} \right)^3.
\label{eq:gaussestimate} 
\end{equation}
The cubic exponent in Eq.(\ref{eq:exmon})
 and Eq.(\ref{eq:gaussestimate}) demonstrates 
that there is a volume effect:
for every condensed boson a minimal volume is required 
that is proportional to $a_x^3$ or $v^3$.

\section{The maximum occupation number for a trapped atom}
For the derivation of Eq.(\ref{eq:gaussestimate})
we assumed Gaussian wave functions.
The wave function of an atom in an harmonic trap will closely resemble this form
for the center-of-mass coordinate.
The intrinsic wave function will not be a Gaussian.
Still we can derive a useful estimate of the MON using a Gaussian approximation.
To set the case we consider hydrogen atoms in a harmonic trap.
Let us define the electron overlap function $f({\bf r}-{\bf r}')$ as
\begin{eqnarray}
    f({\bf r}-{\bf r}') & \equiv &  \int \psi_{int}^{\dagger}({\bf r}-{\bf r}'') 
                                      \psi_{int}({\bf r}''-{\bf r}') d {\bf r}'' 
\nonumber \\ &=&
     \langle \psi_{int} | e^{({\bf r}-{\bf r}')\cdot{\bf \nabla}} 
                        | \psi_{int} \rangle,
\label{eq:foverlap}
\end{eqnarray}
where $\psi_{int}$ denotes the internal wave function of the atom.
The function $f({\bf r})$ is equivalent to the Fourier transform 
of the momentum distribution.
If we can assume that the overlap function is nearly Gaussian,
i.e. if 
\begin{equation}
f({\bf r}) \simeq  e^{-\frac{r^2}{8 v^2}},
\label{eq:fgaussian}
\end{equation}
then the density matrix will still have the form of Eq.(\ref{eq:rhogauss}), 
so that Eq.(\ref{eq:gaussestimate}) applies.
Note that the first order term in a series expansion of $f({\bf r})$
vanishes for any differentiable wave function $ \psi_{int}$, and therefore
Eq.(\ref{eq:fgaussian}) has the correct leading order.
For the $1s$ orbital of the hydrogen atom, the overlap function is given by
$f({\bf r}) = \bar{f}(r/a_0)$, with $a_0$ the atomic Bohr radius and 
\begin{equation}
 \bar{f}(x)=\left(1+x+\frac{x^2}{3}\right) \ e^{-x}.
\end{equation}
Using a series expansion we can write
\begin{equation}
\ln( \bar{f}(x))= -\frac{x^2}{6}+\frac{x^4}{36}-\frac{x^5}{45}+\ldots.
\end{equation}
Keeping only the leading term, we can write
\begin{equation}
f({\bf r}) \simeq e^{-\frac{r^2}{6 a_0^2}}.
\label{eq:fhgaussian}
\end{equation}
This is an approximation, 
but because the next-to-leading order term is of order $r^4$,
we believe that it is good enough 
to allow for a reliable order-of-magnitude estimate of the MON.
The function $F({\bf r})$ in Eq.(\ref{eq:fhgaussian}) 
has the Gaussian form of Eq.(\ref{eq:fgaussian}), with
\begin{equation}
 v \simeq \frac{\sqrt{3}}{2} a_0.
\label{eq:vbohrestimate}
\end{equation}
The MON becomes
\begin{equation}
 \mbox{MON} = \frac{16 W^3 }{3 \sqrt{3} a_0^3}.
\label{eq:hmon}
\end{equation}
Another way to arrive at this result, 
is based on the relation of the displacement operator
$x \cdot \nabla$ to the kinetic energy $T$, valid for s electrons :
\begin{eqnarray}
   \langle \psi_{int}| \left( {\bf r}\cdot{\bf \nabla} \right)^2 
                     | \psi_{int} \rangle 
     &=& - \langle \psi_{int} | \left( {\bf r} \cdot {\bf \hat{p}} 
                   / \hbar \right)^2 | \psi_{int} \rangle 
\nonumber \\ &=&
     - \frac{2 r^2 m_e T }{ 3 \hbar^2},
\end{eqnarray}
with $m_e$ the reduced mass of the electron.
From the {\em virial theorem} we know that the kinetic energy $T$ 
is equal to the electron binding energy $E_b$.
This allows to estimate $v$ as
\begin{equation}
 v \simeq \frac{\sqrt{3} \hbar}{\sqrt{8 m_e E_b}}.
\label{eq:vestimate}
\end{equation}
It is easily checked that Eq.(\ref{eq:vbohrestimate})
and Eq.(\ref{eq:vestimate}) coincide for the hydrogen $1s$ orbital, 
while the latter is a bit more general.

Spin degrees of freedom can be taken into account straightforwardly:
for hydrogen and the alkali atoms,
where the valence electron has one spin-1/2 degree of freedom, 
the estimate of Eq.(\ref{eq:estimate}) for the MON 
has to be multiplied by a factor of 2.

For atoms built up of more than two fermions, the number operator
is no longer equivalent to a fermion pairing operator.
However, we can still apply the above formalism 
by considering one valence electron and the singly ionized atom. 
This makes sense for the alkali atoms,
because the wave function of the first electron out of 
a closed shell has a very long tail
and a small overlap with the other electrons.
In this case, Eq.(\ref{eq:vestimate}) can be used again 
to obtain an order-of-magnitude estimate, 
with $E_b$ replaced by the first ionization energy.
Because of the exchange between valence and core electrons,
it is no longer guaranteed 
that this will result in a strict upper bound for the MON.
However, it is still useful as an order-of-magnitude estimate.
To find a better estimate for the MON in these cases will require
advanced numerical many-body techniques.

Applying Eq.(\ref{eq:vestimate}) to the singly ionized atom 
and the valence electron, we can obtain an estimate for the 
MON from the experimental electron binding energies\cite{Lark77}.
For a trap of width $1 \mu$ we find a MON of the order of $10^{13}$,
for all the alkali atoms.
Present day experiments~\cite{Ande95,Davi95,Brad95,Frie98} 
find occupation numbers of the order of $10^6$ - $10^9$,
which means that they are in a regime of very low densities,
such that Pauli effects due to the constituent fermions 
can safely be neglected.

At higher densities one can expect serious alterations 
in the condensate wavefunction long before the MON is reached, 
due to the interatomic interactions.
However, even at high densities the above formalism remains valid, 
provided that one considers the right one-boson state.
As presented in his famous 1962 paper\cite{Yang62}, 
C.N.~Yang defines the wave function of a Bose condensate
as the eigenvector with the largest eigenvalue of the one-body density matrix.
This Bose wave function can be decomposed as in Eq.(\ref{eq:bosedeco}).
If a condensate would persist even at high densities \cite{Rodr00}, 
its occupation number remains bounded 
by the expression for the MON derived above.
As a matter of fact, nuclear physics might offer an example of this situation.
Recent large-scale shell model calculations \cite{Alha96,Honm01} indicate 
that the valence neutrons of open-shell nuclei tend to form bosonic pairs, 
and that these pairs might have high occupation numbers 
(compared to the number of single-particle states in the system),
that might be very close to their maximum value.

\section{Conclusions}
We conclude that the Pauli principle for the constituent fermions puts 
an upper limit to the occupation number of a composite boson state.
For electron-hole excitons in semiconductors,
this maximum  occupation number can be calculated 
by solving a fermionic generalized pairing problem.
A very accurate approximation to the exact solution shows that
there is a volume dependence in the maximum occupation number:
each exciton consumes a volume that is proportional 
to the extent of the electron-hole intrinsic wave function.
We find a maximum of $V/(4 \pi a_x^3)$ excitons
in a three-dimensional volume $V$ 
( $V /  (\frac{\pi}{2} a_x^2)$ excitons in two dimensions).
The expression also puts an upper limit to the maximum occupation
number of an atomic Bose Einstein condensed state.
A rough estimate based on the electron binding energies, 
leads to values around $10^{13}$ for a trap of width $1\mu$,
much higher than the occupation numbers found in present day experiments.
This means that we can safely look at these condensates 
as truly bosonic condensates, 
though we know that actually they are made up 
of fermionic particles (nucleons and electrons).

\acknowledgments
%
The authors wish to thank J.~Tjon, A.~Dieperink, O.~Scholten, J.~Devreese,
X.~Tempere, L.~Lemmens, K.~Heyde, D.~Dean, M.~Ueda, M.~Honma, T.~Otsuka and D.~Silbar
for the interesting discussions and suggestions,  
and the Fund for Scientific Research - Flanders (Belgium) and the
Research Board of the University of Gent for financial support.



\begin{thebibliography}{0}

\bibitem{Hana77} Hanamura E. and Haug H.,
                 Phys. Rep. {\bf 33} (1977) 209.
\bibitem{Rich64} Richardson R. W. and Sherman N.,
                 Nucl. Phys.{\bf 52} (1964) 221.
\bibitem{Feng98} Feng Pan, Draayer J. P.  and Ormand W. E.,
                 Phys. Lett. B {\bf 422} (1998) 1.
\bibitem{Neck01} Van Neck D., Dewulf Y. and Waroquier M.,
                 Phys. Rev. A {\bf 63} (2001) 062107.
\bibitem{Burg96} Burglin O. and Rowley N., 
                 Nucl. Phys. A  {\bf 602} (1996) 21.
\bibitem{Golu89} Golub G. H. and Van Loan C. F.,
                 {\it Matrix Computations}, The Johns Hopkins University Press, London (1989).
\bibitem{Rice77} Rice T. M.,
                 Solid State Phys. {\bf32} (1977) 1. 
\bibitem{Hens77} Hensel J. C. , Philips T. G. and Thomas G.A.,
                 Solid State Phys. {\bf32} (1977) 88. 
\bibitem{Keld68} Keldysh L. V. and Kozlov A. N.,
                 Zh. Eksp. Teor. Fiz. {\bf54} (1968) 978;
                 [Sov. Phys. JETP {\bf 27}(1968)521].
\bibitem{Comb01} Combescot M. and Tanguy C.,
                 Europhys. Lett. {\bf 55} (2001) 390.
\bibitem{Tang02} Tanguy C.,
                 Phys. Lett. A {\bf292} (2002)285.
\bibitem{Klin95} Klingshirn C. F.,
                 {\it Semiconductor Optics}, Chap. 20 and references therein, 
                 Springer, Berlin (1995).
\bibitem{Lark77} Larkins F. B.,
                 At. Data and Nucl. Data Tables {\bf 20} (1977) 313.
\bibitem{Ande95} Anderson M. H.  {\it et al.},
                 Science {\bf 269} (1995) 198.
\bibitem{Davi95} K. B. Davis {\it et al.},
                 Phys. Rev. Lett. {\bf 75} (1995) 3969.
\bibitem{Brad95} Bradley C. C. {\it et al.},
                 Phys. Rev. Lett. {\bf 75} (1995) 1687.
\bibitem{Frie98} Fried D. G.  {\it et al.},
                 Phys. Rev. Lett. {\bf 81} (1997) 3811.
\bibitem{Yang62} Yang C. N. 
                 Rev. Mod. Phys. {\bf 34} (1962) 694.
\bibitem{Rodr00} Rodriguez B. A. {\it et al.},
                 Int. J. Mod. Phys. B {\bf 14} (2000)71.
\bibitem{Alha96} Alhassid Y., Bertsch G.F., Dean D.J. and Koonin S.E., 
                 Phys. Rev. Lett. {\bf 77} (1996) 1444.
\bibitem{Honm01} Honma M. and Otsuka T.,
                 {\it private communication} (2001).   
\end{thebibliography}
\end{document}